# Tensor Low Rank Modeling and Its Applications in Signal Processing

BABURAJ MADATHIL, Government Engineering College Kozhikode
SAMEERA V MOHD SAGHEER, National Institute of Technology Calicut, Kerala
ABDU RAHIMAN V, Government Engineering College Kozhikode
NISHANTH AUGUSTINE, National Institute of Technology Calicut, Kerala
ANJU JOSE TOM, National Institute of Technology Calicut, Kerala
BAIJU P S, National Institute of Technology Calicut, Kerala
JOBIN FRANCIS, National Institute of Technology Calicut, Kerala
SUDHISH N. GEORGE, National Institute of Technology Calicut, Kerala

Modeling of multidimensional signal using tensor is more convincing than representing it as a collection of matrices. The tensor based approaches can explore the abundant spatial and temporal structures of the mutlidimensional signal. The backbone of this modeling is the mathematical foundations of tensor algebra. The linear transform based tensor algebra furnishes low complex and high performance algebraic structures suitable for the introspection of the multidimensional signal. A comprehensive introduction of the linear transform based tensor algebra is provided from the signal processing viewpoint. The rank of a multidimensional signal is a precious property which gives an insight into the structural aspects of it. All natural multidimensional signals can be approximated to a low rank signal without losing significant information. The low rank approximation is beneficial in many signal processing applications such as denoising, missing sample estimation, resolution enhancement, classification, background estimation, object detection, deweathering, clustering and much more applications. Detailed case study of the ways and means of the low rank modeling in the above said signal processing applications are also presented.

## 1 INTRODUCTION

For the past few decades, multidimensional signals are extensively used for different purposes such as information exchange, creating web content, conducting lectures and product promotions through Internet. Daily, billions of multidimensional signals are exchanged through internet and similar means. The invasion of smartphones into human life is the main reason for huge volume of information exchange. As consequence, the demand for faster and efficient processing algorithms for multidimensional signal is increasing on daily basis. In general, the processing of multidimensional signal is a complex task due to many reasons. First

Authors' addresses: Baburaj Madathil, baburajmadathil@gmail.com, Government Engineering College Kozhikode, P.O. Westhill, Kozhikode, Kerala, 673005; Sameera V Mohd Sagheer, National Institute of Technology Calicut, NIT Campus Post Office, Kozhikode, Kerala, 673601, sameeravm@gmail.com; Abdu Rahiman V, Government Engineering College Kozhikode, P.O. Westhill, Kozhikode, Kerala, 673005, vkarahim@gmail.com; Nishanth Augustine, National Institute of Technology Calicut, NIT Campus Post Office, Kozhikode, Kerala, 673601, nishanthaugustine@gmail.com; Anju Jose Tom, National Institute of Technology Calicut, NIT Campus Post Office, Kozhikode, Kerala, 673601, anjujosetom@gmail.com; Baiju P S, National Institute of Technology Calicut, NIT Campus Post Office, Kozhikode, Kerala, 673601, baijupstvm@gmail.com; Jobin Francis, National Institute of Technology Calicut, NIT Campus Post Office, Kozhikode, Kerala, 673601, jobinkapyarumalayil@gmail.com; Sudhish N. George, National Institute of Technology Calicut, NIT Campus Post Office, Kozhikode, Kerala, 673601, sudhish@nitc.ac.in.

of all, its enormous size distinguishes the multidimensional signal from other signals. Since the size of the signal is huge, the computational complexity also rises to a large extent. Therefore, computationally efficient schemes must be developed to deal with huge size of the data. Moreover, multidimensional signals possess rich structural properties. Accurate modeling of these signals is essential while developing signal processing methods. The existing one/two dimensional algebra is not sufficient for performing this task. In these cases, the multidimensional linear algebra can be employed in the signal processing schemes. The low rank modeling is one of the key ingredient of signal processing tasks [7, 11, 16, 24, 27, 41, 46, 64] . It allows us to approximate higher dimensional signal to a signal living in a low dimensional space.

### 1.1 Motivation

For various signal processing applications, the underlying low rank nature of signals is widely used [11, 16, 64]. For example, in an acquisition system, sensor output may be characterized by a set of closely correlated signals. The ensemble of such signals can be considered as a two dimensional low rank signal. A better perspective of low rank approximation can be illustrated with an image. *Figure*: 1 shows the low rank approximation of an image. When rank of the image is varied from one to five, it is evident that perception of the image is enhanced greatly. Even at a rank of five, it is possible to identify great Joseph Fourier. The signal can be of large rank, but it can be approximated to a low rank signal without much loss of information. The natural images and video signals inherently posses such a low rank structure which can be efficiently utilized as prior for recovering the original data from its corrupted versions or similar applications [16].

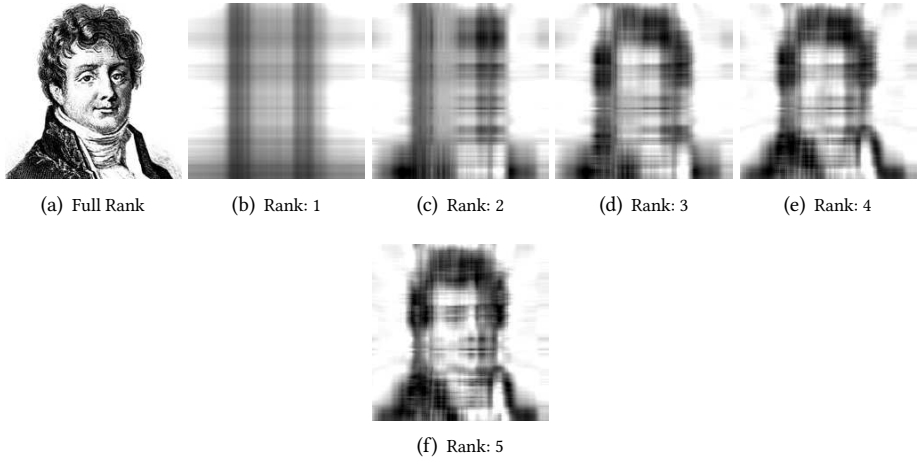

(a) Full Rank  (b) Rank: 1  (c) Rank: 2  (d) Rank: 3  (e) Rank: 4

(f) Rank: 5

**Fig. 1.** Low rank approximation of the image; Joseph Fourier (Full Rank: 256)

Let $\mathbf{M} \in \mathbb{R}^{m \times n}$ be a matrix, then the rank of $\mathbf{M}$ is defined as the dimension of the vector space spanned by its columns or rows. The rank varies when the signal is affected by impairments [16]. Suppose the singular value decomposition (SVD) of $M$ is given by [16],

$$\mathbf{M} = \mathbf{U}\mathbf{S}\mathbf{V}^T \tag{1}$$

where, $\mathbf{U}$ and $\mathbf{V}$ are respectively the matrices containing left and right singular vectors and $\mathbf{S}$ is the singular value matrix. The singular value vector of $\mathbf{M}$ is defined as, $\sigma = diag(\mathbf{S})$. Rank of a two dimensional signal can also be defined as the number of non-zero singular values [16]. Mathematically,

$$Rank(\mathbf{M}) = \|\sigma\|_0 \tag{2}$$

where, $\|.\|_0$ indicates $l_0$ norm. The impact of rank on signal impairments is shown in *Figure*: 2. A low rank signal and its singular values are shown respectively in *Figure*: 2(a) and *Figure*: 2(b). It is observed that, the singular values exhibit a fast decay characteristics [47]. The impaired signals are shown in *Figure*: *2(c)* and *Figure*: *2(e)* which are affected by noise and lost samples respectively. When impairments are introduced into the signal, its rank increases. New significant singular values are appeared in the decomposition as shown in the *Figure*: *2(d)* and *Figure*: *2(f)*. When signal is affected by impairments, the low rank prior can be employed for recovering the original information. The newly created singular values due to impairment can be disregard by selective thresholding [16] for obtaining the original low rank signal. Hence, multidimensional signals affected by degradations can be effectively restored by extracting the underlying low rank representation.

However, the non-convexity nature of $l_0$ norm limits its application [11]. Hence, the $l_0$ norm can be replaced with its convex approximation called $l_1$ norm which makes the problem tractable [16]. The $l_1$ norm of singular value vector is called the nuclear norm of a matrix [47] and is defined as,

$$\|\mathbf{M}\|_* = \|\boldsymbol{\sigma}\|_1 \tag{3}$$

Since, the rank of corrupted signal is large, its nuclear norm will have large value than that of the original signal. The corrupted signal can be constrained with nuclear norm to obtain its original low rank approximation and hence, nuclear norm minimization is considered as the solution for low rank recovery problem. The low rank recovery model can be formulated as [11],

$$\arg\min_{\mathbf{X}} Rank(\mathbf{X}) \quad \text{such that} \quad \mathbf{X} = \mathbf{Y} \tag{4}$$

where, $Y \in \mathbb{R}^{m \times n}$ is the impaired signal. Since $Rank()$ is non-convex, it can be replaced with nuclear norm for converting the problem into convex form as given below [64].

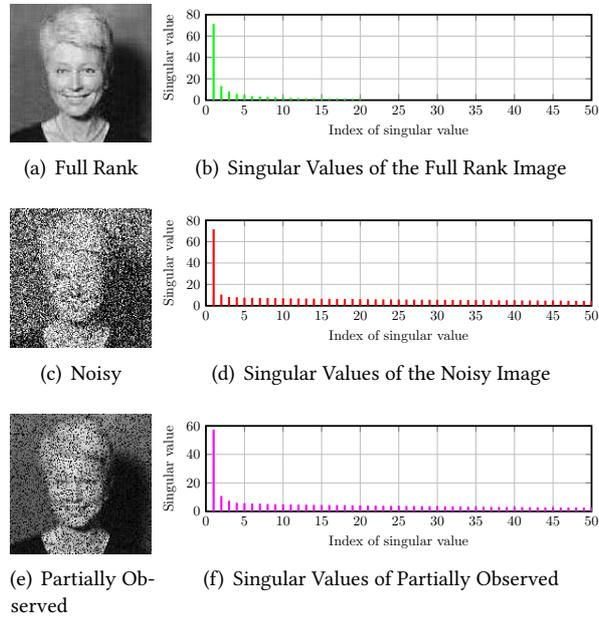

(a) Full Rank  (b) Singular Values of the Full Rank Image

(c) Noisy  (d) Singular Values of the Noisy Image

(e) Partially Observed  (f) Singular Values of Partially Observed

**Fig. 2.** Impact of rank on signal impairments

$$\arg\min_{\mathbf{X}} \|\mathbf{X}\|_* \quad \text{such that} \quad \frac{1}{2}\|\mathbf{X} - \mathbf{Y}\|_F^2 \tag{5}$$

The above problem can be solved using singular value thresholding [64]. These concepts can also be extended to higher dimensional signals modeled as tensors [27]. Tensor may be considered as a multidimensional array. Tensor framework [27] can be utilized to formulate the low rank representation of multidimensional signals. When multidimensional signals are affected by impairments, the rank minimization problem based recovery schemes can be designed. Hence, rank of a signal plays an important role in the eradication of signal impairments and nuclear norm constrained recovery models can be effectively utilized for signal recovery.

In summary, natural multidimensional signals such as video and images possess low rank structure [11, 12, 16, 17, 28, 64]. Even though, signal lies in a high dimensional space, it can be approximated to live in a relatively low dimensional space by exploiting low rank property. The low rank approximation is one of the approaches used to obtain the low dimensional estimate of multilinear data with least error. Use of nuclear norm in low rank approximation of matrices is well established and widely used in image recovery [18, 20].

### 1.2 The Development of Tensor Frameworks

Higher dimensional signal can be considered as ensemble of matrices. However, spatial as well as temporal redundancy exist in multilinear signals. If these matrices are considered individually and build its low rank approximation, then the temporal correlation cannot be effectively utilized. This thought leads to the development of tensor algebraic frameworks for low rank approximation of multilinear signals (tensors). [7, 13, 27, 41].

Low rank factorization is the means to disclose the structure of tensors as extensions of matrix SVD notions. There has been significant analysis in this direction and a number of frameworks along with algebra and rank definitions exist. One among the primitive and elementary algebraic factorization is CANonical DECOMPosition PARAllel FACtors (CANDECOMP / PARAFAC) [13] factorization in which the tensor is disintegrated as a finite sum of rank-1 tensors. Additionally, tensor rank is defined as the minimum number of rank-1 tensors for accurate decomposition. Although this appears to be conceptually simpler, the problem is heavily ill-posed because the rank is unknown in prior essentially. On the contrary, Tucker [7] decomposition is more likely with higher order Principal Component Analysis (PCA), with a core tensor to measure the interaction with components along with transformed matrices on each mode. It can also be approximated as the sum of outer products of vectors similar to CANDECOMP / PARAFAC (CP) [13].

Another important direction is based on a group theoretic approach, where the multilinear structure is unravelled by forming group-rings along the tensor fibres [7, 27, 41]. The benefit of such an approach over the existing methods is that the resulting algebra and analysis is very similar to that of matrix algebra and analysis. Linear transforms like DFT is used to exploit the temporal correlation existing in the signals. Computational cost of these techniques are less compared with CP or Tucker because they do not use any optimization techniques to perform the factorization. Third order tensor factorization techniques were studied by *Martin et al.* [41] in a recursive manner for a general $p$-order tensor. Nevertheless, multiplying blocks of block circulant matrices is costly in terms of computational speed and memory. Encouraged from the principle of circulant matrix diagonalization, *Martin et al.* developed a straightforward solution for higher order tensor multiplication using Fast Fourier Transform (FFT) [41]. Motivated by these works, *Kilmer et al.* [27] proposed an improved tensor framework based on circulant algebra. They proposed tensor- singular value decomposition (t-SVD) based on Fourier transform and proposed the idea of tensor multirank.

### 1.3 Notations and Tensor Algebra

A tensor is a multilinear structure [16, 42] in $\mathbb{R}^{n_1 \times n_2 \times \cdots n_N}$. A vector is first order tensor, a matrix is second order tensor and multilinear data of order three or above are called higher order tensors. A third order tensor is shown in *Figure*: 3. A more practical case is illustrated in *Figure*: 4 in which we consider a video sequence as a third order tensor. Basic notations of the tensor is summarized in *Table*: A.

Data can be extracted from a tensor is many ways in the form of matrix or vector. The data extraction methods are summarized in *Table*: B. A slice of a tensor in a 2D section which is defined by all but two indices [42]. The horizontal, lateral and frontal slices of the third order tensor are shown in *Figure*: 5. For a 3-way tensor $\mathcal{X}$, $k^{th}$ horizontal, lateral and frontal slices are given by $\mathcal{X}(k,:,:)$, $\mathcal{X}(:,k,:)$ and $\mathcal{X}(:,:,k)$ respectively. A fiber of a tensor is a 1D section defined by fixing all indices but one. The mode-1, mode-2 and mode-3 fibers of a third order tensor are shown in *Figure*: 6. The fibers $\mathcal{X}(:,i,j)$, $\mathcal{X}(i,:,j)$ and $\mathcal{X}(i,j,:)$ denote mode-1, mode-2 and mode-3 fibers respectively.

The circulant algebra based tensor framework [27, 42] is the recently developed one for tensor low rank representation. The spatio-temporal structures of a multidimensional signal is effectively utilized by transforming third mode fibers using an appropriate transform [27, 42]. The well known Discrete Fourier Transform (DFT) is used to transform the tensor along the third mode. The $\mathcal{X}_f = \mathit{fft}(\mathcal{X}, 3)$ denotes $\mathit{fft}(.)$ along the third dimension. In order to illustrate the effectiveness of tensor framework [27, 42], it defines five block based operations namely, $bcirc(.), bvec(.), bvfold(.), bdiag(.)$ and $bdfold(.)$. For a tensor $\mathcal{X} \in \mathbb{R}^{n_1 \times n_2 \times n_3}$, five block-based operations are summarized in *Table*: C. The tensor framework defined various operations on tensor which is listed in *Table*: D. The tensor product (t-product) allows to multiply two tensors through circular convolution. However, the operation can be carried out efficiently in transform domain. The tensor SVD called t-SVD [27] is the backbone of low rank representation. The t-SVD allows us to decompose a tensor, $\mathcal{X}$ into t-product of three tensors $\mathcal{X} = \mathcal{U} * \Sigma * \mathcal{V}^T$, where, $\mathcal{U}$ and $\mathcal{V}$ are left and right unitary singular tensors, and $\Sigma$ is the f-diagonal tensor containing singular values of $\mathcal{X}$. The singular value tensor $\Sigma$ determines the low rank characteristics of the it. The t-SVD is illustrated in *Figure*: 7. The rank of a multidimensional signal (multirank) is defined as a vector whose each element represents the rank of the corresponding frontal slices of $\mathcal{X}_f$. The nuclear norm of the tensor is also defined based on t-SVD as given in *Table*: D. The block matrix based representation can also be used for computing these parameters. Block circulant representation of $\mathcal{X}$ preserves the spatial relationship between entries, and $\|bcirc(\mathcal{X})\|_*$ measures the rank of $bcirc(\mathcal{X})$ by comparing every row and every column of frontal slices over the third dimension which exploits the spatio-temporal information of a tensor much deeper than the monotonous matrix nuclear norm of certain unfolding [22]. Certain signal processing applications requires rearrangement of tensor slices and these operations are called twist and squeeze as illustrated in *Figure*: 8.

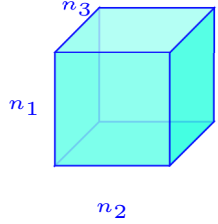

Fig. 3. A three dimensional Tensor

| Notation | Description |
| --- | --- |
| Euler script | Tensors, e.g. $\mathcal{X}$ |
| Bold upper case letters | Matrices, e.g. $\mathbf{M}$ |
| Bold lower case letters | Vectors, e.g. $\mathbf{v}$ |
| Lower case letters | Scalers |
| $\mathcal{X}$ | A third order tensor in $\mathbb{R}^{n_1 \times n_2 \times n_3}$ |
| $\mathcal{Y}$ | A third order tensor in $\mathbb{R}^{n_2 \times n_4 \times n_3}$ |
| $\mathcal{Z}$ | A third order tensor in $\mathbb{R}^{n_1 \times n_4 \times n_3}$ |
| $\mathcal{I}$ | A third order tensor in $\mathbb{R}^{n \times n \times n_3}$ |
| $\mathcal{U}$ | A third order tensor in $\mathbb{R}^{n \times n \times n_3}$ |
| $\mathcal{V}$ | A third order tensor in $\mathbb{R}^{n_2 \times n_2 \times n_3}$ |
| $\Sigma$ | A third order tensor in $\mathbb{R}^{n_1 \times n_2 \times n_3}$ |

Table A. Basic notations

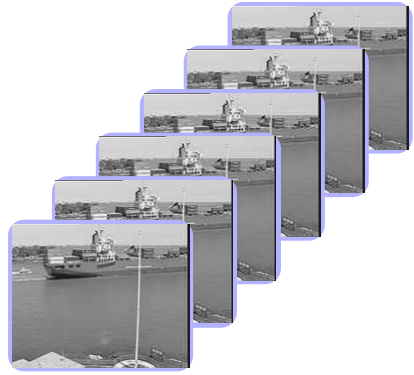

Fig. 4. A video sequence as three dimensional tensor

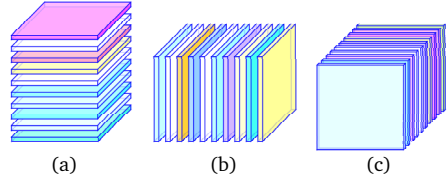

Fig. 5. (a) Horizontal (b) Lateral (c) Frontal slices of a third order tensor

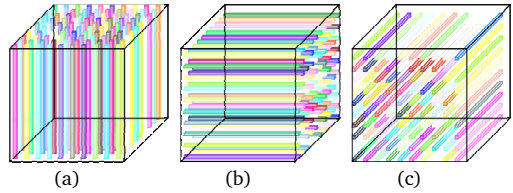

Fig. 6. (a) mode-1 fiber (b) mode-2 fiber (c) mode-3 fiber

## 1.4 Tensor Low Rank Approximation

Similar to matrix low rank approximation problem discussed in *Section*: 1.1, we can formulate tensor low rank approximation problem as [22, 60],

$$\arg\min_{\mathcal{X}} \ \|\mathcal{X}\|_\circledast \quad \text{such that} \quad \frac{1}{2}\|\mathcal{X} - \mathcal{Y}\|_F^2 \tag{6}$$

where, $\mathcal{Y} \in \mathbb{R}^{n_1 \times n_2 \times n_3}$ is the observed tensor. The above problem can be solved using tensor framework using tensor singular value thresholding [22, 36]. Let $\mathcal{Y} = \mathcal{U} * \Sigma * \mathcal{V}^T$ be the singular value decomposition of $\mathcal{Y}$ and $\tau$ is the given threshold, then the tensor singular value thresholding operator $\mathfrak{D}_\tau(.)$ is defined as [22],

$$\mathfrak{D}_\tau[\mathcal{Y}] = \mathcal{U} * C_\tau(\Sigma) * \mathcal{V}^T \qquad (7)$$

$$C_\tau(\Sigma) = \Sigma * \mathcal{J}$$

where, $\mathcal{J}$ is an $n_1 \times n_1 \times n_3$ $f$-diagonal tensor whose diagonal element in the Fourier domain is,

$$\mathcal{J}_f(i,i,k) = \left(1 - \frac{\tau}{\Sigma_f(i,i,k)}\right)_+, \ 1 \le i \le n_1, \ 1 \le k \le n_3 \qquad (8)$$

where, $\Sigma_f$ is the singular values of the tensor $\mathcal{Y}$ in Fourier domain. The optimum value of $\mathcal{X}$ is given by,

$$\hat{\mathcal{X}} = \mathfrak{D}_\tau[\mathcal{Y}] \qquad (9)$$

| Notation | Description |
| --- | --- |
| $\mathcal{X}(k,:,:)$ | Horizontal slices of the tensor $\mathcal{X}$ |
| $\mathcal{X}(:,k,:)$ | Lateral slices of the tensor $\mathcal{X}$ |
| $\mathcal{X}(:,:,k)$ | Frontal slices of the tensor $\mathcal{X}$ |
| $\mathcal{X}(:,i,j)$ | Mode-1 fiber of the tensor $\mathcal{X}$ |
| $\mathcal{X}(i,:,j)$ | Mode-2 fiber of the tensor $\mathcal{X}$ |
| $\mathcal{X}(i,j,:)$ | Mode-3 fiber of the tensor $\mathcal{X}$ |
| $\mathcal{X}^{(k)}$ | $k^{th}$ frontal slice of tensor $\mathcal{X}$ |

**Table B.** Data extractions methods from tensor in the form of vector or matrix

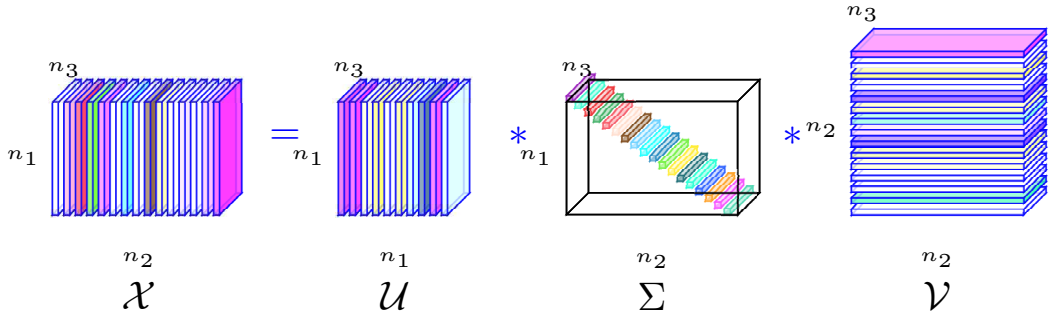

**Fig. 7.** Tensor Singular Value Decomposition

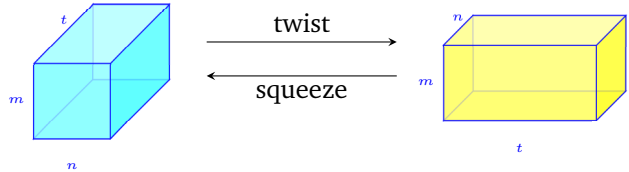

**Fig. 8.** Twist and squeeze operation of $m \times n \times t$ tensor

| Operation | Description |
|---|---|
| Mode-$l$ unfolding | A matrix whose columns are mode-$l$ fibers, $X_{(l)} = unfold_l(X)$ |
| Mode-$l$ folding | Reverse of unfold, $fold_l(X_{(l)}) = X$ |
| Block circulant matrix $bcirc(X)$ | $\begin{bmatrix} X^{(1)} & X^{(n_3)} & \ldots & X^{(2)} \\ X^{(2)} & X^{(1)} & \ldots & X^{(3)} \\ \vdots & \vdots & \vdots & \vdots \\ X^{(n_3)} & X^{(n_3-1)} & \ldots & X^{(1)} \end{bmatrix}$ |
| Block vector matrix $bvec(X)$ | $\begin{bmatrix} X^{(1)} \\ X^{(2)} \\ \vdots \\ X^{(n_3)} \end{bmatrix}$ |
| Fold block vector matrix $bvfold(.)$ | Reverse operation of $bvec(.)$, $X = bvfold(bvec(X))$ |
| Block diagonal matrix $bdiag(X)$ | $\begin{bmatrix} X^{(1)} & & \\ & \ddots & \\ & & X^{(n_3)} \end{bmatrix}$ |
| Fold block diagonal matrix $bdfold(.)$ | Reverse operation of $bdiag(.)$, $X = bdfold(bdiag(X))$ |
| Twisted tensor $\vec{X}$ | Lateral slices of $\vec{X}(:, k, :) = X^{(k)}$ |
| $squeeze\left(\vec{X}\right)$ | Reverse of twist operation, $X = squeeze\left(\vec{X}\right)$ |

**Table C.** Operations involving rearrangement of slices and fibers of a tensor

In the next sections, development and applications of the tensor low rank approximation model in various signal processing scenarios are discussed.

## 2 MULTIDIMENSIONAL SIGNAL COMPLECTION

Signal completion intends to retrieve the original signal from its incomplete measurements. Practical measurements may be deteriorated arbitrarily with respect to signal position and magnitude. Noise corruption and occlusions turn practical image and video signals quite noisy and annoying even with recent advancements. Certain regions of image or frame segments might be invisible for users due to scratches, occlusions or errors while data conversion/communication

Tensor Low Rank Modeling and Its Applications in Signal Processing9Tensor Low Rank Modeling and Its Applications in Signal Processing 9

| Operation | Description |
|---|---|
| Transformed tensor $\mathcal{X}_f$ | Computed by taking $fft(.)$ along the third mode fibers of tensor $\mathcal{X}$ |
| Frobenius Norm $\|\mathcal{X}\|_F$ | $\sqrt{\sum_{i,j,k} \mathcal{X}(i,j,k)^2}$ |
| $l_1$-norm $\|\mathcal{X}\|_1$ | $\sum_{i,j,k} |\mathcal{X}(i,j,k)|$ |
| Tensor Product $\mathcal{Z} = \mathcal{X} * \mathcal{Y}$ | $\mathcal{Z} = bvfold(bcir(\mathcal{X})bvec(\mathcal{Y}))$ |
| Fast Tensor Product $\mathcal{Z}\mathcal{X} * \mathcal{Y}$ | $\mathcal{Z}_f^{(k)} = \mathcal{X}_f^{(k)} \mathcal{Y}_f^{(k)}, \quad k = 1, \ldots, n_3$ |
| Transpose $\mathcal{X}^T$ | Transpose each frontal slice of $\mathcal{X}$ and then reverse the order of the transposed frontal slices 2 through $n_3$, element of $\mathbb{R}^{n_2 \times n_1 \times n_3}$ |
| Identity Tensor $\mathcal{I}$ | First frontal slice is an $n_1 \times n_1$ identity matrix and all other frontal slices are zero matrices. |
| Unitary Tensor $\mathcal{U}$ | $\mathcal{U}^T * \mathcal{U} = \mathcal{U} * \mathcal{U}^T = n\mathcal{I}, \quad n \in \mathbb{R}$ |
| f-diagonal Tensor | Frontal slices of the tensor are diagonal matrices |
| Inner Product $\langle \mathcal{X}, \mathcal{Y} \rangle$ | $trace(bdiag(\mathcal{X})^H bdiag(\mathcal{Y}))$ |
| Tensor Singular Value Decomposition (t-SVD) | $\mathcal{X} = \mathcal{U} * \Sigma * \mathcal{V}^T$, $\mathcal{U}$ and $\mathcal{V}$ are left and right unitary singular tensors, and $\Sigma$ is the f-diagonal tensor containing singular values of $\mathcal{X}$. |
| $multirank(\mathcal{X})$ | A vector in $\mathbb{R}^{n_3}$ with $i^{th}$ element equal to the rank of $i^{th}$ frontal slice of $\mathcal{X}_f, i = 1, \ldots n_3$ |
| Nuclear Norm $\|\mathcal{X}\|_{\circledast}$ | $\sum_{k=1}^{n_3} \sum_{i=1}^{min(n_1,n_2)} |\Sigma_f(i,i,k)|$ |

**Table D.** Elementary operations of the linear transform (DFT) based tensor framework

## 2.1 Low Rank Matrix Completion

The matrix completions aims to restore an image which suffers from lost samples. In an image completion problem, suppose $\mathbf{M} \in \mathbb{R}^{m \times n}$ and $\mathbf{X} \in \mathbb{R}^{m \times n}$ correspond to degraded and recovered versions of an image respectively. Restoration from limited samples is modelled as [22],

$$\min_{\mathbf{X}} \quad rank(\mathbf{X}) \quad \text{such that} \quad \mathcal{P}_\Omega(\mathbf{X}) = \mathcal{P}_\Omega(\mathbf{M}) \tag{10}$$

Where, $\Omega$ indicates the index set of the reliable pixels and $\mathcal{P}_\Omega(.)$ is the orthogonal projection onto the span of matrices with non-zero entries restricted to $\Omega$. Low rank signal recovery broadly handles reconstructing from partial knowledge by regularizing the rank. Though $rank(.)$ of a matrix is well defined, rank minimization is an NP hard problem. Thus, the minimization is to run with the best matching convex surrogate namely, nuclear norm [18] and the convex model is given by,

$$\min_{\mathbf{X}} \quad \|\mathbf{X}\|_* \qquad \text{such that} \quad \mathcal{P}_\Omega(\mathbf{X}) = \mathcal{P}_\Omega(\mathbf{M}) \tag{11}$$

Use of nuclear norm for optimization is favourable in two ways, first of all convexity seeks global minimum irrespective of initial conditions. Secondly, nuclear norm is proven to be the tightest convex surrogate of rank [4]. The above problem is called Low Rank Matrix Completion (LRMC) [16, 18, 61, 64] and it is employed in image completion, video completion [64], medical image reconstruction [64] etc.

## 2.2 Low Rank Tensor Completion

The low rank matrix completion method is inefficient for retrieving multilinear data because of the matrix nuclear norm cannot capture the spatial redundancy and temporal redundancy together [36]. Nowadays, a very large volume of multilinear data is created in the form of video, hyperspectral images, multispectral images, magnetic resonance images etc. Since matrix completion is not an effective method to process these signals, it is considered to treat them as tensors and processes signals by taking full advantage of the methods implemented in tensor algebra [10, 27]. A major hurdle in this area is that rank of a multilinear data is not well defined [13]. In summary, the low rank tensor completion problem is defined as completing an observed tensor, $\mathcal{M} \in \mathbb{R}^{n_1 \times n_2 \times n_3}$ from available entries, $\Omega$,

$$\min_{\mathcal{X}} \quad multirank(\mathcal{X}) \qquad \text{such that} \quad \mathcal{P}_\Omega(\mathcal{X}) = \mathcal{P}_\Omega(\mathcal{M}) \tag{12}$$

where, $\mathcal{X} \in \mathbb{R}^{n_1 \times n_2 \times n_3}$ is the estimate of underlying low rank tensor and $\mathcal{P}_\Omega(.)$ is the orthogonal projection onto the span of tensors with non-zero entries restricted to $\Omega$. Due to nonconvexity of $multirank(.)$, the above problem is ill-posed and convex model is given by [36, 63],

$$\min_{\mathcal{X}} \quad \|\mathcal{X}\|_\circledast \qquad \text{such that} \quad \mathcal{P}_\Omega(\mathcal{X}) = \mathcal{P}_\Omega(\mathcal{M}) \tag{13}$$

The above low rank tensor completion model [22, 29, 30, 61] can be applied effectively to multilinear data completion problems such as video completion, hyperspectral image completion, medical image completion etc. A typical video completion application is illustrated in *Figure*: 9. Frames of partially observed and estimated versions of the bus [52] video sequence is shown in the *Figure*: 9.

## 3 MULTIDIMENSIONAL SIGNAL DECOMPOSITION

Consider the case, where a low rank matrix/tensor is perturbed by a sparse matrix/tensor. In general, the sparse matrix/tensor can have entries of arbitrary magnitude. The objective is to decompose the signal to estimate the low rank and sparse components and these types of situations transpire in many signal processing problems such as denoising, restoration, inpainting etc.

## 3.1 Low Rank Matrix Decomposition

Low rank matrix reconstruction from corrupted matrix is an enriching research topic. An image or video corrupted by gross errors can be recovered with high accuracy by regularizing its rank [4, 16, 28, 36, 64]. A solution can be reached by simultaneously minimizing rank and sparsity of the signal. But rank operator is non-convex operator and its closest convex approximation is nuclear norm as mentioned [4]. Suppose a signal $\acute{\mathbf{L}} \in \mathbb{R}^{M \times N}$ is corrupted by gross errors and it is required to recover the signal from the available observations, $\mathbf{M} \in \mathbb{R}^{M \times N}$. Then, the observed signal, $\mathbf{M}$ can be decomposed into low rank and sparse components as $\mathbf{M} = \acute{\mathbf{L}} + \acute{\mathbf{S}}$, where $\acute{\mathbf{L}}$ is the low rank component and $\acute{\mathbf{S}}$ is the sparse component. The decomposition is illustrated in *Figure*: 10. If $\acute{\mathbf{L}}$ is low rank and satisfies incoherence conditions and $\acute{\mathbf{S}}$ is sufficiently sparse, then $\acute{\mathbf{L}}$ and $\acute{\mathbf{S}}$ can be recovered from $\mathbf{M}$ with high probability by solving the following optimization problem [4],

$$\min_{\mathbf{L},\mathbf{S}} \quad \|\mathbf{L}\|_* + \lambda \|\mathbf{S}\|_1 \qquad \text{such that} \quad \mathbf{M} = \mathbf{L} + \mathbf{S} \tag{14}$$

where, $\lambda$ is regularization parameter. The above problem is called Robust Principal Component Analysis (RPCA) [4] and it is employed in image restoration, medical image reconstruction, moving object detection etc. [64].

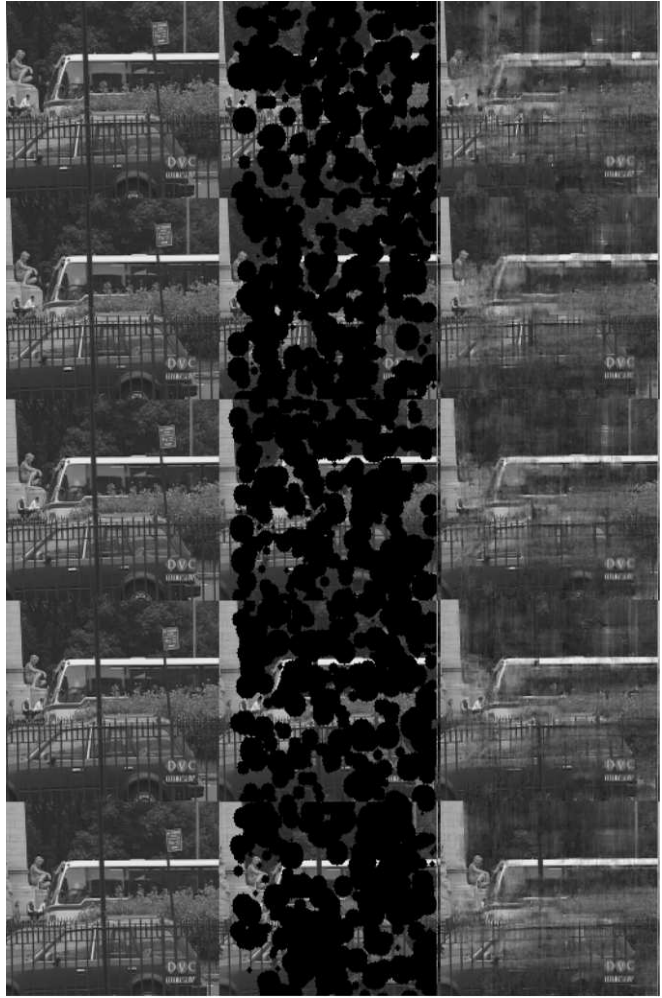

**Fig. 9.** Illustration of tensor completion for video signal missing sample estimation: Original signal, partially observed signal having 70% sample loss and estimated signal are shown first, second and third columns of the figure respectively.

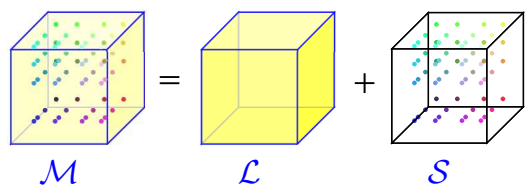

**Fig. 10.** Tensor decomposition model for signal denoising

### 3.2 Low Rank Tensor Decomposition

The problems in which a higher dimensional signal perturbed by sparse multidimensional signal can be solved by observing the signals as a tensor. The developed multilinear algebra of tensor can be effectively employed to recover the degraded tensor from gross errors. The low rank tensor decomposition problem is defined as decomposing an observed multidimensional data, $\mathcal{M}$ which is corrupted by gross errors, into a low rank component $\acute{\mathcal{L}}$ and sparse component $\acute{\mathcal{S}}$ so that $M = \acute{\mathcal{L}} + \acute{\mathcal{S}}$. The Tensor Robust Principal Component Analysis (TRPCA) [36] problem is stated as,

$$\min_{\mathcal{L},\mathcal{S}} \quad \|\mathcal{L}\|_{\circledast} + \lambda \|\mathcal{S}\|_1 \qquad \text{subject to} \quad \mathcal{M} = \mathcal{L} + \mathcal{S} \tag{15}$$

where $\|.\|_{\circledast}$ is the tensor nuclear norm and $\lambda$ is the regularization parameter. The above problem is successively employed in video restoration [36], hyperspectral anomaly detection [55] etc. A typical video densoising application is illustrated in *Figure*: 11. Frames of noisy and denoised versions of the bus [52] video sequence is shown in the *Figure*: 11.

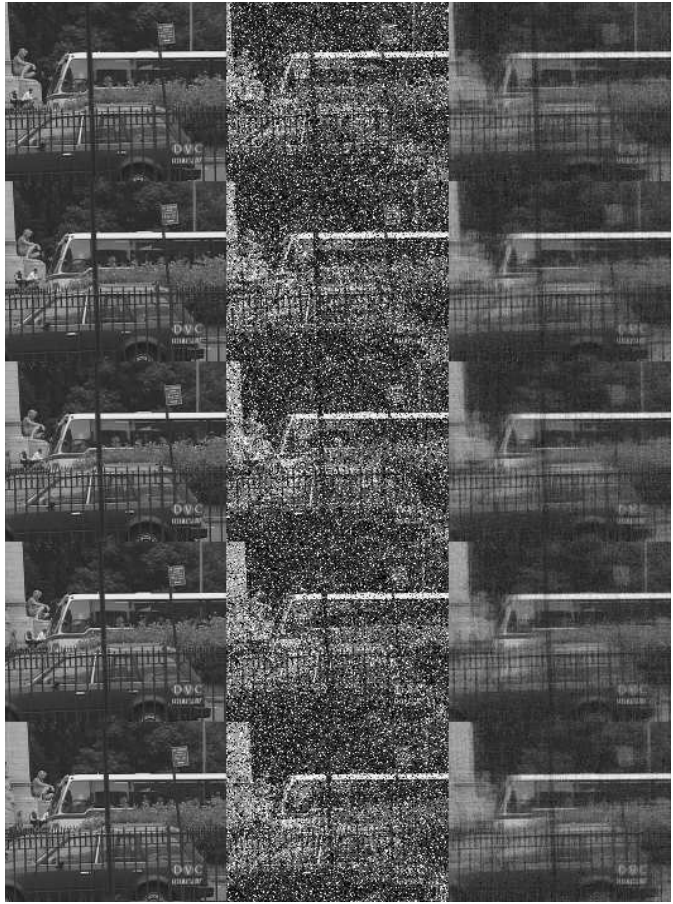

**Fig. 11.** Illustration of tensor decomposition for video denoising: Original signal, noisy signal with $\sigma^2 = 0.2$ and estimated signal are shown first, second and third columns of the figure respectively.

## 4 LOW RANK TENSOR MODELS FOR MEDICAL IMAGE DENOISING

The advancements in image processing techniques have created a wave of change in the field of medical diagnostics. The role of laboratory based diagnosis is reduced to a great extent and medical imaging is the new trend in the diagnosis of diseases. The presence of noise or artifacts which may be introduced during the process of acquisition, further processing and/or transmission of images can lead to the misinterpretation of diseases. This preample makes denoising of medical images an unavoidable pre-processing step in disease diagnosis using medical images. The requisites of a satisfactory medical image denoising algorithm can be listed as i) preservation of edges and other finer details, ii) maintenance of structural similarity, iii) nonappearance of artifacts and iv) low complexity of operations.

Conventional medical imaging modalities include Ultrasound (US), Magnetic Resonance (MR), Computed Tomography (CT) and Positron Emission Tomography (PET) images. However, with the rapid growth in technology, $3D$ and $4D$ imaging technologies have become the new trend in medical imaging techniques. The classical image denoising algorithms aim to segregate the noise free image from a corrupted data. The literature presents a plethora of techniques which accomplish the denoising process by representing the data using matrix algebra, where, each frame is considered as a matrix. However, as mentioned earlier the use of matrix algebra can result in loss of structural information. Also, the direct extension of matrix based method spoils the temporal correlation present in the medical images which affects the denoising performance adversely. This is because the matrix nuclear norm defined in *Eq*: (1.1) cannot capture the spatial redundancy and temporal redundancy simultaneously [59]. The concept of tensors introduced earlier can be exploited for representing multidimensional data. The different frames of such a medical image have strong correlations. Moreover, from the spatial viewpoint, its intensity is almost similar to that of its neighbourhood. The medical image is treated as a tensor by exploiting the spatial and temporal redundancy present in the data. Also, the correlation between the different frames can be modeled using low rank tensor penalty. Since noise characteristics of different medical imaging modalities are different, the tensor low rank approximation techniques mentioned in Section *Section*: 1.4 cannot be directly applied for denoising of the medical images. Recently a few works were proposed for denoising of Magnetic Resonance (MR) images [15, 25, 26] and and Computer Tomography (CT) images [50] by exploiting the low rank nature of the data. Hawazin *et al.* introduced a tensor based denoising for MR images [25]. MR images are corrupted with signal dependent Rician noise. The Variance Stabilization Technique (VST) introduced by Foi *et al.* [14] is applied to remove the signal dependency of the noise. Low rank tensor approximation is eventually utilized to model the noisy data after VST as an $n^{th}$ order tensor. The estimate of denoised image $\hat{\mathcal{X}}$ for a given noisy image $\mathcal{Y} \in \mathbb{R}^{n_1 \times n_2 \times n_3}$ and a noise-free image $\mathcal{X} \in \mathbb{R}^{n_1 \times n_2 \times n_3}$, is obtained by minimizing the rank of the noisy observation. The rank minimization problem is NP hard and non-convex in nature. Thus, the rank is replaced by its closest convex surrogate, the Tensor Nuclear Norm (TNN). The TNN penalizes all the singular values equally. It does not consider the fact that the higher singular values contain greater information and hence needs to be preserved more. Therefore, the TNN is replaced by the Weighted Tensor Nuclear Norm (WTNN) [38] which assigns a weighting factor during the thresholding of the singular values. Hence, the denoising problem can be reformulated as,

$$\hat{\mathcal{X}} = \min_{\mathcal{X}} \| \mathcal{Y} - \mathcal{X} \|_F^2 + \lambda \| \mathcal{X} \|_{WTNN} \tag{16}$$

Finally, the denoised image in the original domain is obtained by performing inverse VST. For better clarity, a graphical illustration of the above concept is presented in Fig. 12. In addition, the qualitative results obtained for this method is demonstrated in Fig. 13.

## 5 LOW RANK MODELS FOR IMAGE SUPER RESOLUTION

Image super resolution is the signal processing technique to recover high resolution (HR) images from already captured low resolution (LR) observations [45]. A simplified image formation model is given in Fig. 14 and it can be mathematically denoted as, [45, 57],

$$\mathbf{Y}_k = \mathbf{H}_k \mathbf{X} + \mathbf{N}_k, \text{for } k = 1 \cdots K \tag{17}$$

where, $\mathbf{X}$ is the HR image to synthesize, $\mathbf{H}_k$ denotes the combined effect of blurring and down sampling on $k^{th}$ observation $\mathbf{Y}_k$, $\mathbf{N}_k$ is the noise effects on $k^{th}$ image and $K$ is the number of LR observations. For single image super resolution (SISR), $K = 1$. In an ideal image acquisition system the desired HR image $\mathbf{X}$ is the expected output, but due to the errors in the practical acquisition process, the observed image is a degraded version of the desired image. Super resolution is an inverse signal recovery problem that computes **X** from $\mathbf{Y}_k$.

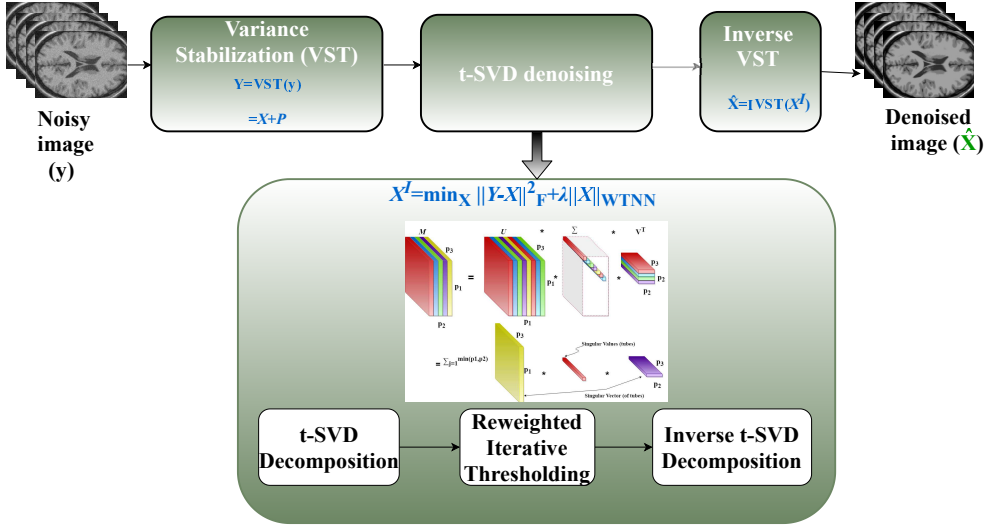

**Fig. 12.**  Basic block diagram of the tensor based MR image denoising

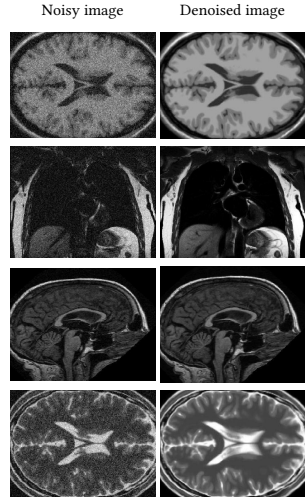

**Fig. 13.**  Denoised results of real MR experiment (OASIS database [40])

Super resolution is formulated as a regularized optimization problem with appropriate constraints as [44],

$$\mathbf{X} = \arg\min_{\mathbf{X}} \frac{1}{2}\|\mathbf{HX} - \mathbf{Y}\|_F^2 + \lambda \mathbf{R}(\mathbf{X}) \tag{18}$$

where, $\lambda$ is the regularization parameter and **R (X)** is the regularization function. Low rank models and total variation functions are commonly used as regularization function in such inverse problems to stabilize the inversion process [57] [8] [34] [51]. Regularization term represents image model with prior knowledge about the desired image. Nuclear norm regularization is a smoothness prior to enforce low rank property of natural images on the reconstructed image. Total variation (TV), which is an edge preserving prior model, is included to avoid over smoothing of edges [51]. Overall objective function is given as,

$$\mathbf{X} = \arg\min_{\mathbf{X}} \lambda_1 \|\mathbf{X}\|_* + \lambda_2 \|\mathbf{X}\|_{TV}$$
$$\text{Such that } \mathbf{HX} = \mathbf{Y} \quad (19)$$

where, $\lambda_1$ and $\lambda_2$ are the manually selected regularization parameters to balance the contributions of each term to the objective function. $\mathbf{H}$ is created for given blur kernel and magnification factor. Figure 15 gives the result of super resolution for a magnification factor of 4 and *Gaussian* blur kernel.

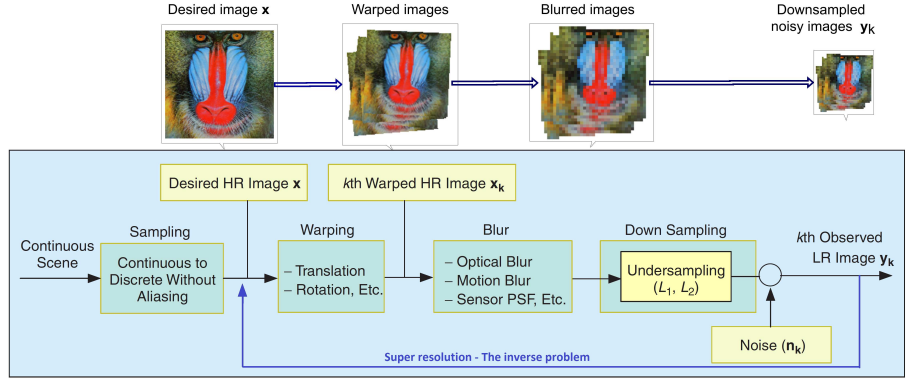

**Fig. 14.** A simplified image acquisition model for image super resolution.

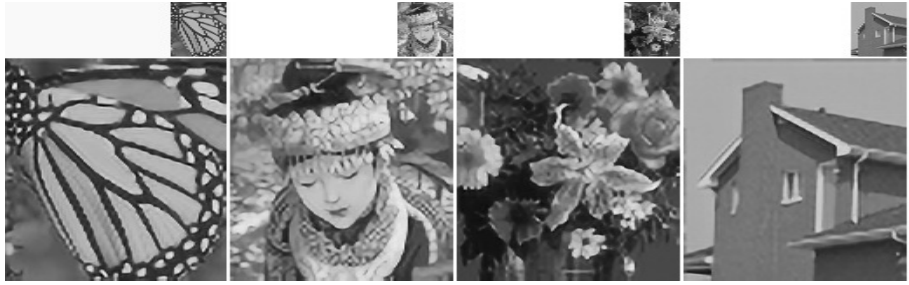

**Fig. 15.** Results of super resolution using low rank and total variation based approach for a magnification factor of 4. Top row is the input image and the bottom row is the super resolved output.

## 6 HYPERSPECTRAL IMAGE DENOISING VIA TENSOR LOW RANK RECOVERY

Hyperspectral images (HSI) are widely used in remote sensing (RS) for material detection and identification in many areas, such as geography, agriculture, military etc. HS images are acquired by using hundreds of contiguous spectral bands and hence provides numerous spectral information [9]. Since, the signal received by spectrometers at the satellite or on the ground are affected by the interference from the atmosphere and from the instrument itself, HS images are encountered by various types of noises [58]. The presence of these noises degrades the visual quality of the HSI data and adversely affects the performance of subsequent processes such as, image classification, fusion, unmixing etc. Hence, noise reduction is an important pre-processing step in hyperspectral image analysis [19].

Generally, during acquisition, the HS images are corrupted by a mixture of different noises such as Gaussian noise, impulse noise, dead lines, stripes etc. The Gaussian noise is due to the atmospheric absorption and instrumental noise. HS images may suffer from the impulse noise due to the unpredictable fault of some sensors [2]. The reason for stripes and dead line noises in HS image is the physical damage of detectors [49]. Until now, various methods have been proposed for HS image denoising [56]. Earlier methods extend the traditional 1-D or 2-D denoising methods to HS image band by band [62]. The performance of these methods are not satisfactory, as the spatial and spectral ccorrelation are not simultaneously considered [19]. Hence these methods are only capable of removing either spatial or spectral noise.

Efficient HS image denoising can only be achieved by considering spatial and spectral information jointly [19]. This can be fulfilled by taking the HSI as a whole entity. Recently, many tensor algebra based low rank approximation methods were proposed in the context of multi-frame image processing, which utilized the multi-linear algebra. Since HS image can be treated as a multi-frame image, tensor algebra can be effectively used for the direct analysis of HS images [32]. The use of tensors reduces the computational overheads considerably. Any tensor decomposition model (PARAFAC, TUCKER3 and t-SVD) can be used for HSI denoising [43] but t-SVD is preferred since it is less computationally complex.

Since, the low rank tensor approximation (LRTA) [48] method only uses the spectral correlation and local spatial consistency is not taken into proper consideration, the results may suffer from spatial distortions. A solution to this problem is to incorporate tensor total variation regularization (TVR) [21] approach along with LRTA. TVR excellently preserves local spatial consistency in the underlying signal. Normally, the bands of an HS image exhibit strong spectral correlation and if considered each band as a matrix, they exhibits relatively strong spatial correlation. This spatial and spectral correlation can be modeled by a low-rank tensor penalty. In addition to this, in spatial domain, the intensity of each voxel seems to almost equal to those in its neighborhood, and the same is true from the spectral domain point of view also. This local spatial and spectral smoothness property can be described using 3-D tensor total variation [6]. Thus, by combining low-rank penalty and 3-D tensor total variation, complete utilization of spectral and spatial correlations of HSI can be achieved and hence an efficient HSI denoising method can be implemented.

As mentioned earlier, HS images are corrupted by additive Gaussian noise and sparse noises like impulse noise, dead lines and stripes. The observed HS image which is corrupted by mixed noise can be written as,

$$\mathcal{H} = \mathcal{L} + \mathcal{S} + \mathcal{N} \tag{20}$$

where,

- $\mathcal{H} \in R^{m \times n \times o}$ is the observed noisy HS image having $o$ bands and $mn$ pixels in each band.
- $\mathcal{L} \in R^{m \times n \times o}$ is the underlying low rank tensor (i.e. clean HS image) in the observed HS image.
- $\mathcal{S} \in R^{m \times n \times o}$ is the sparse noise.
- $\mathcal{N} \in R^{m \times n \times o}$ is the additive Gaussian noise.

The HS image mixed denoising process aims at restoring the clean HS image $\mathcal{L}$ from the corrupted data $\mathcal{H}$. In LRTA, the denoising problem can be stated as,

$$\arg\min_{\lim} \mathcal{L}_{i,j}, \mathcal{S}_{i,j}, \mathcal{N}_{i,j} \left[ \left\| \mathcal{L}_{i,j} \right\|_{\circledast} + \lambda \left\| \mathcal{S}_{i,j} \right\|_1 + \tau \left\| \mathcal{N}_{i,j} \right\|_F^2 \right] \\ s.t \;\; \mathcal{H}_{i,j} = \mathcal{L}_{i,j} + \mathcal{S}_{i,j} + \mathcal{N}_{i,j} \tag{21}$$

Since, the LRTA [31] method only uses the spectral constraint and local spatial consistency is not taken into proper consideration, some spatial distortions may occur. This problem can be solved by incorporating total variation denoising (TVD) [6] approach along with LRTA. As mentioned earlier, TVD excellently preserves local spatial consistency in the underlying signal.

So, by combining LRTA model and total variation minimizaton term, the proposed denoising problem can be modeled as,

$$\arg\min_{\lim} \mathcal{L}_{i,j}, \mathcal{S}_{i,j}, \mathcal{N}_{i,j} \Big[ \left\| \mathcal{L}_{i,j} \right\|_{\circledast} + \lambda \left\| \mathcal{S}_{i,j} \right\|_1 + \tau \left\| \mathcal{N}_{i,j} \right\|_F^2 \\ + \gamma TV(\mathcal{L}_{i,j}) \Big] \\ s.t \;\; \mathcal{H}_{i,j} = \mathcal{L}_{i,j} + \mathcal{S}_{i,j} + \mathcal{N}_{i,j} \tag{22}$$

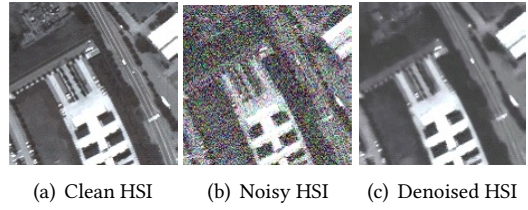

(a) Clean HSI  (b) Noisy HSI  (c) Denoised HSI

Fig. 16. Result of pavia data set

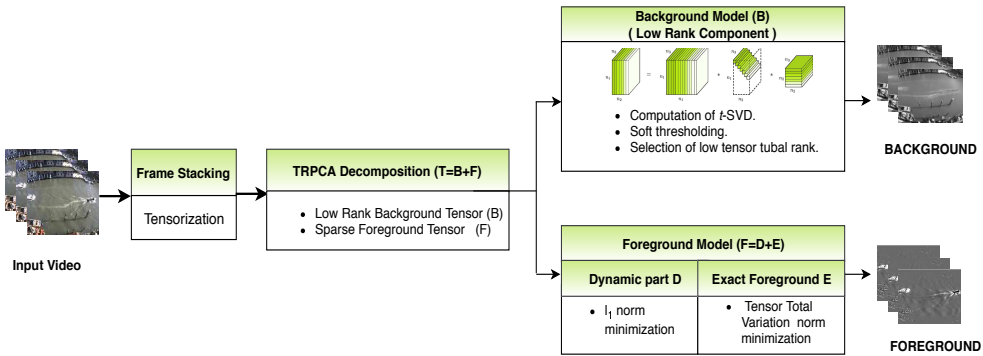

Fig. 17. Basic block diagram of the RPCA based MOD method

Denoised output is obtained by solving above problem using ADMM [3]. The result of Pavia data set corrupted by additive Gaussian noise of variance 0.4 is shown in Fig. 16.

## 7 TENSOR RPCA DECOMPOSITION IN MOVING OBJECT DETECTION APPLICATIONS FOR SURVEILLANCE VIDEOS

The conventional Background Subtraction (BS) and Moving Object Detection (MOD) problems operates on the matrix framework considering each frame as a matrix or vectorizing frames as columns. This form of flattening or resizing results in loss of structural information. In other words, the hidden information can be revealed only if the analysis tool accounts for the multi-way model present in the input data. Alternatively, a 3 way tensor treatment of the video data throughout the implementation ensures efficient utilization of its structural properties. As it is known, there are more than 200 million cameras in daily use globally and in fact, a camera operator can miss almost 90% of the activity on a screen after 20 minutes during video monitoring task. The objective of the classical MOD methods is to segregate a sparse blend of moving objects called foreground from the static portion of structured well-regulated and organized information called the background. In addition, there are some challenging environments usually encountered such as swaying trees, moving water, waves etc. which are themselves moving (known as dynamic background). The background may not be available to use and can always be changed in such cases. So, the background representation model must be more robust and adapted accordingly. Instead of restructuring the 3D data to matrix, RPCA [4] can be extended to the tensor framework [37]. This forms a clarification for the systematic recovery of low rank tensor and gathering sparse data from the superposition of both these components. Consider a video sequence, $\mathcal{T} \in \mathbb{R}^{n_1 \times n_2 \times n_3}$ where, $n_1$ represents the width, $n_2$ gives the height of the video frame and the total number of frames is given by $n_3$. If the short duration video volume has a single channel (R,G, or B/grayscale), then it can be represented as a $3^{rd}$ order tensor which upholds the spatial and temporal features of the video input. The principle of TRPCA essentially complies with this idea such that [4]:

$$\min_{\mathcal{B},\mathcal{F}} \quad rank(\mathcal{B}) + \lambda\|\mathcal{F}\|_0 \qquad (23)$$
$$\text{s.t.} \quad \mathcal{T} = \mathcal{B} + \mathcal{F}$$

where, $\mathcal{B}$ is the structured low rank part constituting video background and $\mathcal{F}$ is the sparse component portion including the dynamic disturbance and this part contains the moving objects too. $\lambda$ is the weight controlling coefficient of the sparse tensor and is chosen as, $\lambda = \frac{1}{\sqrt{max(n_1,n_2)\times n_3}}$ [37]. The tensor tubal rank and $l_0$ norm are not convex quantities so the minimization is done by considering their surrogates, Tensor Nuclear Norm (TNN) and $l_1$ norm. Moreover, Eq. (23) performs well when the background is static (the camera is static, eg: shopping mall). In real life video sequences, locally induced background motion will form dynamic natured backgrounds. So the low tensor tubal rank [37] constraint is not enough. Basically, the background appears with high spatio-temporal correlation whereas foreground has a coherent property which makes it move smoothly in the frames. From this idea, $\mathcal{F}$ can be decomposed into a dynamic portion, $\mathcal{D}$ representing the disturbance arising from emotive nature of background and $\mathcal{E}$ indicating the exact foreground or moving object in the video volume. What differentiates the dynamic background and moving foreground, is the spatio-temporal continuity of the foreground. Thus exact foreground part can be separated if the spatio-temporal continuity is enhanced. In addition to this, the dynamic background is sparser than the exact foreground part and $l_1$ norm minimization keeps it small or ignored. Hence the constrained optimization problem can be formulated as follows [6]:

$$\min_{\mathcal{B},\mathcal{F},\mathcal{D},\mathcal{E}} \quad \|\mathcal{B}\|_\circledast + \lambda_1\|F\|_1 + \lambda_2\|D\|_1 + \lambda_3\|\Omega(E)\| \qquad (24)$$
$$\text{s.t.} \quad \mathcal{T} = \mathcal{B} + \mathcal{F} \quad \mathcal{F} = \mathcal{D} + \mathcal{E}$$

where, $\|.\|_\circledast$ denotes the Tensor Nuclear Norm, $\lambda_1, \lambda_2$ and $\lambda_3$ are the weights for balancing Eq. (27) and $\Omega(.)$ regularises the foreground to make it coherent in the spatial direction and smooth in the temporal direction. This can be made possible by incorporating Tensor Total Variation norm [39] as the regularization term.

$$\min_{\mathcal{B},\mathcal{F},\mathcal{D},\mathcal{E}} \quad \|\mathcal{B}\|_\circledast + \lambda_1\|F\|_1 + \lambda_2\|D\|_1 + \lambda_3\|\mathcal{E}\|_{TTV} \qquad (25)$$
$$\text{s.t.} \quad \mathcal{T} = \mathcal{B} + \mathcal{F} \quad \mathcal{F} = \mathcal{D} + \mathcal{E}$$

$\|\mathcal{E}\|_{TTV}$ is the Tensor Total Variation Norm. Eq. 25 can be solved using ALM with Alternating direction

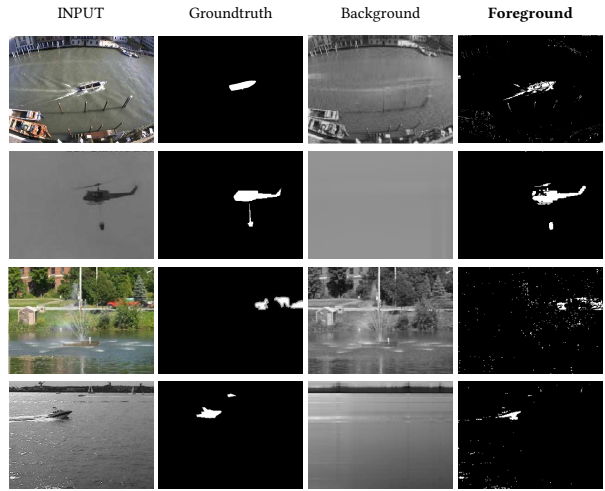

**Fig. 18.** 1-4 columns respectively illustrate Input Video, Ground Truth, Background and Foreground obtained using Proposed method ; Top row : $MAR$ sequence, Second row: UCSD-$chopper$, Third row: CD.net-$fountain$02, Bottom row: UCSD-$boats$. For the foreground masks, white represents correctly detected foreground.

approach. For better understanding, the basic block diagram of the RPCA based MOD method is illustrated in Fig. 17. In addition, the qualitative results obtained for this method is demonstrated in Fig. 19.

## 8 LOW RANK TENSOR DECOMPOSITION BASED RAIN REMOVAL SCHEME FOR SURVEILLANCE VIDEOS

Generally, outdoor computer vision systems are designed for video surveillance applications such as vehicle tracking, crowd analysis, automatic number plate detection, military applications, animal monitoring in forestry, etc. Such vision systems can provide better performance in clement climatic conditions. However, inclement weather conditions such as rain can cause degradation in visual quality of outdoor surveillance videos. This degradation will severely affect the performance of the vision systems and hence the accuracy may become poor. Therefore, it is necessary to remove the undesirable visual artifacts caused by rain on outdoor videos so that the vision systems can achieve better performance and greater accuracy. Existing rain removal methods can be categorized into two: Learning based and reconstruction based methods. Learning based methods require huge data set and time consuming training phase. However, reconstruction based methods use single data to get rain free results. One of the leading and emerging reconstruction based techniques is low rank tensor decomposition approach. In [23], rainy video is modelled as,

$$O = \mathcal{B} + \mathcal{R} \quad (26)$$

where, $O, \mathcal{B}$ and $\mathcal{R} \in \mathbb{R}^{n_1 \times n_2 \times n_3}$ are input rainy video, rain free video and rain streak video respectively. Here, rain removal problem can be considered as separating $\mathcal{B}$ and $\mathcal{R}$ from $O$. Since clean video exhibits inherent low rank nature, rank minimization will help to extract $\mathcal{B}$ from $O$ [5]. Unlike background video, rain streaks are sparser in nature [1]. Hence, its sparsity can be boosted by minimizing $l_1$ norm of $\mathcal{R}$. Moreover, derivatives of rain streaks along the direction rain fall are more sparser than those of clean background. Hence, the smoothness along the rain direction can be enhanced by employing unidirectional TV regularization along the direction of ran streaks [35]. Due to the presence of rain streaks, smoothness of clean video along horizontal as well as temporal directions has been lost. Hence by applying TV minimization of clean video along horizontal and temporal directions, smoothness of clean video can be enhanced. Thus the rain removal problem can be formulated as constrained optimization problem as,

$$\min_{\mathcal{B},\mathcal{R}} \|\mathcal{B}\|_\circledast + \lambda_1 \|\mathcal{R}\|_1 + \lambda_2 \|\nabla_y \mathcal{R}\|_1 + \lambda_3 \|\nabla_x \mathcal{B}\|_1$$
$$+ \lambda_4 \|\nabla_t \mathcal{B}\|_1 \quad (27)$$
$$\text{s.t.} \quad O = \mathcal{B} + \mathcal{R}$$

The experimental results obtained for this method are shown in Fig. 19.

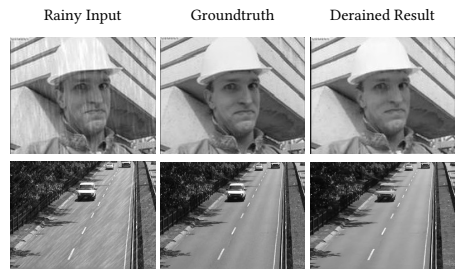

**Fig. 19.** 1-3 columns respectively represent Rainy Input, Ground Truth and Derained videos.

## 9 CLUSTERING OF IMAGING DATA USING LOW RANK APPROACH

Clustering is a data analysis technique widely used in natural groupings of large data sets. In scientific data analysis, a parametric model is required to characterize a given set of data. The intention of such a modelling is to extract some feature of interest embedded in the data. So clustering has attracted remarkable attention in many areas such as data mining, machine learning, computer vision and pattern recognition. However a given data cannot be well addressed by a single subspace. A reasonable way of representation is to consider the data as lying near a union of subspaces (UoS) and the clustering can be based on their closeness to individual subspaces[33]. Low rank representation (LRR) is a well-known Dimensionality Reduction (DR) technique for subspace clustering, which projects the higher dimensional data into a low dimensional space.

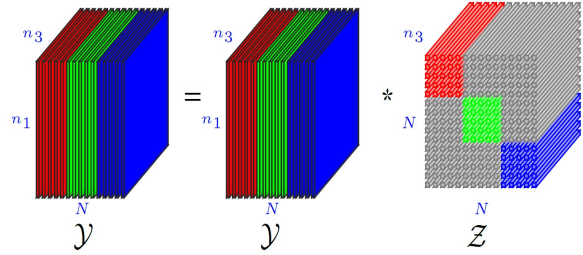

**Fig. 20.** Figure illustrates self expressiveness property: an image in a free submodule is represented as the linear combination of other images belonging to the same submodule.

In this section we discuss the UoFS model (Union of Free Sub module)[53] for clustering of two dimensional data such as images. Consider the set of $N$ images that may belong to distinct categories with each image represented by $\{Y_j \in \mathbb{R}^{n_1 \times n_3}\}_{j=1}^{N}$. Then, the challenge is to segment the entire set of images into different, say $L$ clusters. This particular problem can be modeled as considering a tensor $\mathcal{Y} \in \mathbb{R}^{n_1 \times N \times n_3}$ and the images to be clustered are stacked as the lateral slices of the same tensor $\mathcal{Y}$. Incorporating the algebra concepts into tensors, the analogue of a vector subspace is a free submodule. The assumption is that those images belong to a union of $L$ free submodules (UoFS) and the goal is to segment those images into individual clusters with each cluster corresponding to a submodule.

It is supposed that data points in low dimensional space preserves self- expressiveness property [54]. According to this, one data point can be represented by a linear combination of other data points. i.e. an image belonging to a free submodule can be represented as $t$-linear combination of other images belonging to the same free submodule. It means that there exist a tensor, $\mathcal{Z} \in \mathbb{R}^{N \times N \times n_3}$ such that $\mathcal{Y} \approx \mathcal{Y} \times \mathcal{Z}$.

In addition, images belonging to one submodule are not supposed to be represented by images belonging to another submodules. Then, the tensor $\mathcal{Z}$ is obtained that are permutations of $f$-block diagonal structure. Once such a representation tensor $Z^*$ is obtained, an affinity matrix W can be made with each $(i,j)^{th}$ entry $W_{ij}$ is given by[53]

$$w_{ij} = \| \mathcal{Z}^*(i,j,:) \|_F + \| \mathcal{Z}^*(j,i,:) \|_F \tag{28}$$

Spectral clustering is implemented as a last step to obtain the final clustering. This model is a unified framework of submodule clustering with low rank representation. Therefore the obtained $\mathcal{Z}$ is not only expected to approximate the data tensor ($\mathcal{Y} \approx \mathcal{Y} \times \mathcal{Z}$); possessing a low multirank as well. It can be achieved by minimizing its $TNN$ [53]. To impose the f-block diagonal structure on $\mathcal{Z}$, an assumption can be followed that images belonging to different free submodules must have lower correlations. This particular intuition is captured in terms of a dissimilarity matrix $M \in [0,1]^{N \times N}$. Similarity and dissimilarity measures are referred to as the measures of proximity. Dissimilarity measure returns a value of 1 if the objects being considered are similar and a value of 1 if those objects are dissimilar. A single objective function is then formulated which integrates low rank, $f$-block diagonal structure and the dissimilarity measure. Finally the problem can be stated as,

$$\min_{\mathcal{Z}} \|\mathcal{Z}\|_{\circledast} + \lambda_1 \sum_{k=1}^{n_3} \| M \odot Z \|_1 + \lambda_2 \| \mathcal{Y} - \mathcal{Y} \times \mathcal{Z} \|_F^2 \tag{29}$$

In the above expression, $\odot$ denotes element wise multiplication and $\lambda_1$ and $\lambda_2$ are the regularization parameters. To solve the this expression, variable splitting can be applied and rearranging the above equation as[53]

$$\min_{C, Q, \mathcal{Z}} \|\mathcal{Z}\|_{\circledast} + \lambda_1 \sum_{k=1}^{n_3} \| M \odot Q^{(k)} \|_1 + \lambda_2 \| \mathcal{Y} - \mathcal{Y} \times \mathcal{Z} \|_F^2 \tag{30}$$
$$\text{s.t.} \mathcal{Z} = C \text{ and } \mathcal{Z} = Q$$

The above optimization problem can be solved using using ADMM [3] approach.

## 10 CONCLUSIONS & FUTURE DIRECTIONS

In today's world, modern devices are generating multidimensional data in huge volumes. These multidimensional data possesses spatio-temporal correlations and abundant structural properties. Dedicated algorithms need to be developed for efficient processing of these signals. Conventional matrix based approaches are insufficient to gather the structural properties of the multidimensional signals. We have shown that the tensor based approaches are the best tools to explore multidimensional data. The tensor based approaches allow us to fully exploit the redundancies existing in all directions of the signal. The developing tensor algebra opens up enormous possibilities in the multidimensional signal processing area.

Low rankness is an important characteristic of a natural multidimensional signal that can be effectively utilized as a prior in signal processing applications. Rank constrained signal recovery models can be developed by employing mathematical building blocks presented in the tensor linear algebra. We have discussed these mathematical building blocks in detail. The discussion continued to explore the role of tensor low rank models in diverse signal processing applications. In all the applications presented, low rankness of the signal represented as tensor is utilized to improve the efficacy of the algorithms.

In the future, methods can be developed to work without linearity assumption. Natural problems might be characterized by nonlinearities. Therefore, nonlinear models can deliver better results to natural signal recovery problems. In addition to that, the incoherence condition of tensor decomposition model can be further investigated to accommodate weaker assumptions. The low rank models can be further explored to find new applications areas. The computational complexity of algorithms can be reduced by analyzing the possibilities of parallel processing.

## REFERENCES

[1] Alaa E Abdel-Hakim. 2014. A novel approach for rain removal from videos using low-rank recovery. In *2014 5th International Conference on Intelligent Systems, Modelling and Simulation*. IEEE, 351–356.

[2] Hemant Kumar Aggarwal and Angshul Majumdar. 2015. Mixed Gaussian and impulse denoising of hyperspectral images. In *2015 IEEE International Geoscience and Remote Sensing Symposium (IGARSS)*. IEEE, 429–432.

[3] Stephen Boyd, Neal Parikh, Eric Chu, Borja Peleato, Jonathan Eckstein, et al. 2011. Distributed optimization and statistical learning via the alternating direction method of multipliers. *Foundations and Trends® in Machine learning* 3, 1 (2011), 1–122.

[4] Emmanuel J Candès, Xiaodong Li, Yi Ma, and John Wright. 2011. Robust principal component analysis? *Journal of the ACM (JACM)* 58, 3 (2011), 11.

[5] Wenfei Cao, Yao Wang, Jian Sun, Deyu Meng, Can Yang, Andrzej Cichocki, and Zongben Xu. 2016. Total variation regularized tensor RPCA for background subtraction from compressive measurements. *IEEE Transactions on Image Processing* 25, 9 (2016), 4075–4090.

[6] Xiaochun Cao, Liang Yang, and Xiaojie Guo. 2015. Total variation regularized RPCA for irregularly moving object detection under dynamic background. *IEEE transactions on cybernetics* 46, 4 (2015), 1014–1027.

[7] J Douglas Carroll and Jih-Jie Chang. 1970. Analysis of individual differences in multidimensional scaling via an N-way generalization of "Eckart-Young" decomposition. *Psychometrika* 35, 3 (1970), 283–319.

[8] Antonin Chambolle. 2004. An algorithm for total variation minimization and applications. *Journal of Mathematical imaging and vision* 20, 1-2 (2004), 89–97.

[9] Chein-I Chang. 2007. *Hyperspectral data exploitation: theory and applications*. John Wiley & Sons.

[10] Andrzej Cichocki, Danilo Mandic, Lieven De Lathauwer, Guoxu Zhou, Qibin Zhao, Cesar Caiafa, and Huy Anh Phan. 2015. Tensor decompositions for signal processing applications: From two-way to multiway component analysis. *IEEE Signal Processing Magazine* 32, 2 (2015), 145–163.

[11] Pierre Comon. 2014. Tensors: a brief introduction. *IEEE Signal Processing Magazine* 31, 3 (2014), 44–53.

[12] Lieven De Lathauwer. 2009. A survey of tensor methods. In *2009 IEEE International Symposium on Circuits and Systems*. IEEE, 2773–2776.

[13] Lieven De Lathauwer, Bart De Moor, and Joos Vandewalle. 2000. A multilinear singular value decomposition. *SIAM journal on Matrix Analysis and Applications* 21, 4 (2000), 1253–1278.

[14] Alessandro Foi. 2011. Noise estimation and removal in MR imaging: The variance-stabilization approach. In *2011 IEEE International symposium on biomedical imaging: from nano to macro*. IEEE, 1809–1814.

[15] Ying Fu and Weisheng Dong. 2016. 3d magnetic resonance image denoising using low-rank tensor approximation. *Neurocomputing* 195 (2016), 30–39.

[16] Donald Goldfarb and Zhiwei Qin. 2014. Robust low-rank tensor recovery: Models and algorithms. *SIAM J. Matrix Anal. Appl.* 35, 1 (2014), 225–253.

[17] Lars Grasedyck, Daniel Kressner, and Christine Tobler. 2013. A literature survey of low-rank tensor approximation techniques. *GAMM-Mitteilungen* 36, 1 (2013), 53–78.

[18] Qiang Guo, Caiming Zhang, Yunfeng Zhang, and Hui Liu. 2015. An efficient SVD-based method for image denoising. *IEEE transactions on Circuits and Systems for Video Technology* 26, 5 (2015), 868–880.

[19] Xian Guo, Xin Huang, Liangpei Zhang, and Lefei Zhang. 2013. Hyperspectral image noise reduction based on rank-1 tensor decomposition. *ISPRS journal of photogrammetry and remote sensing* 83 (2013), 50–63.

[20] Wei He, Hongyan Zhang, Liangpei Zhang, and Huanfeng Shen. 2015. Total-variation-regularized low-rank matrix factorization for hyperspectral image restoration. *IEEE transactions on geoscience and remote sensing* 54, 1 (2015), 178–188.

[21] Wei He, Hongyan Zhang, Liangpei Zhang, and Huanfeng Shen. 2015. Total-variation-regularized low-rank matrix factorization for hyperspectral image restoration. *IEEE transactions on geoscience and remote sensing* 54, 1 (2015), 178–188.

[22] Wenrui Hu, Dacheng Tao, Wensheng Zhang, Yuan Xie, and Yehui Yang. 2016. The twist tensor nuclear norm for video completion. *IEEE transactions on neural networks and learning systems* 28, 12 (2016), 2961–2973.

[23] Tai-Xiang Jiang, Ting-Zhu Huang, Xi-Le Zhao, Liang-Jian Deng, and Yao Wang. 2017. A novel tensor-based video rain streaks removal approach via utilizing discriminatively intrinsic priors. In *Proceedings of the ieee conference on computer vision and pattern recognition*. 4057–4066.

[24] Eric Kernfeld, Misha Kilmer, and Shuchin Aeron. 2015. Tensor–tensor products with invertible linear transforms. *Linear Algebra Appl.* 485 (2015), 545–570.

[25] Hawazin S Khaleel, Sameera V Mohd Sagheer, M Baburaj, and Sudhish N George. 2018. Denoising of Rician corrupted 3D magnetic resonance images using tensor-SVD. *Biomedical Signal Processing and Control* 44 (2018), 82–95.

[26] Hawazin S Khaleel, Sameera V Mohd Sagheer, M Baburaj, and Sudhish N George. 2018. Denoising of Volumetric MR Image Using Low-Rank Approximation on Tensor SVD Framework. In *Proceedings of 2nd International Conference on Computer Vision & Image Processing*. Springer, 371–383.

[27] Misha E Kilmer, Karen Braman, Ning Hao, and Randy C Hoover. 2013. Third-order tensors as operators on matrices: A theoretical and computational framework with applications in imaging. *SIAM J. Matrix Anal. Appl.* 34, 1 (2013), 148–172.

[28] Tamara G Kolda and Brett W Bader. 2009. Tensor decompositions and applications. *SIAM review* 51, 3 (2009), 455–500.

[29] Daniel Kressner, Michael Steinlechner, and Bart Vandereycken. 2014. Low-rank tensor completion by Riemannian optimization. *BIT Numerical Mathematics* 54, 2 (2014), 447–468.

[30] Akshay Krishnamurthy and Aarti Singh. 2013. Low-rank matrix and tensor completion via adaptive sampling. In *Advances in Neural Information Processing Systems*. 836–844.

[31] Chang Li, Yong Ma, Jun Huang, Xiaoguang Mei, and Jiayi Ma. 2015. Hyperspectral image denoising using the robust low-rank tensor recovery. *JOSA A* 32, 9 (2015), 1604–1612.

[32] Tao Lin and Salah Bourennane. 2013. Survey of hyperspectral image denoising methods based on tensor decompositions. *EURASIP journal on Advances in Signal Processing* 2013, 1 (2013), 186.

[33] Guangcan Liu, Zhouchen Lin, and Yong Yu. 2010. Robust subspace segmentation by low-rank representation.. In *ICML*, Vol. 1. 8.

[34] Ji Liu, Przemyslaw Musialski, Peter Wonka, and Jieping Ye. 2012. Tensor completion for estimating missing values in visual data. *IEEE transactions on pattern analysis and machine intelligence* 35, 1 (2012), 208–220.

[35] Ji Liu, Przemyslaw Musialski, Peter Wonka, and Jieping Ye. 2012. Tensor completion for estimating missing values in visual data. *IEEE transactions on pattern analysis and machine intelligence* 35, 1 (2012), 208–220.

[36] Canyi Lu, Jiashi Feng, Yudong Chen, Wei Liu, Zhouchen Lin, and Shuicheng Yan. 2016. Tensor robust principal component analysis: Exact recovery of corrupted low-rank tensors via convex optimization. In *Proceedings of the IEEE conference on computer vision and pattern recognition*. 5249–5257.

[37] Canyi Lu, Jiashi Feng, Yudong Chen, Wei Liu, Zhouchen Lin, and Shuicheng Yan. 2016. Tensor robust principal component analysis: Exact recovery of corrupted low-rank tensors via convex optimization. In *Proceedings of the IEEE conference on computer vision and pattern recognition*. 5249–5257.

[38] Baburaj Madathil and Sudhish N George. 2018. Twist tensor total variation regularized-reweighted nuclear norm based tensor completion for video missing area recovery. *Information Sciences* 423 (2018), 376–397.

[39] Baburaj Madathil and Sudhish N George. 2018. Twist tensor total variation regularized-reweighted nuclear norm based tensor completion for video missing area recovery. *Information Sciences* 423 (2018), 376–397.

[40] Daniel S Marcus, Tracy H Wang, Jamie Parker, John G Csernansky, John C Morris, and Randy L Buckner. 2007. Open Access Series of Imaging Studies (OASIS): cross-sectional MRI data in young, middle aged, nondemented, and demented older adults. *Journal of cognitive neuroscience* 19, 9 (2007), 1498–1507.

[41] Carla D Martin, Richard Shafer, and Betsy LaRue. 2013. An order $p$ tensor factorization with applications in imaging. *SIAM Journal on Scientific Computing* 35, 1 (2013), A474–A490.

[42] Carla D Martin, Richard Shafer, and Betsy LaRue. 2013. An order $p$ tensor factorization with applications in imaging. *SIAM Journal on Scientific Computing* 35, 1 (2013), A474–A490.

[43] Damien Muti, Salah Bourennane, and Julien Marot. 2008. Lower-rank tensor approximation and multiway filtering. *SIAM journal on Matrix Analysis and Applications* 30, 3 (2008), 1172–1204.

[44] Kamal Nasrollahi and Thomas B Moeslund. 2014. Super-resolution: a comprehensive survey. *Machine vision and applications* 25, 6 (2014), 1423–1468.

[45] Sung Cheol Park, Min Kyu Park, and Moon Gi Kang. 2003. Super-resolution image reconstruction: a technical overview. *IEEE signal processing magazine* 20, 3 (2003), 21–36.

[46] Yigang Peng, Jinli Suo, Qionghai Dai, and Wenli Xu. 2014. Reweighted low-rank matrix recovery and its application in image restoration. *IEEE transactions on cybernetics* 44, 12 (2014), 2418–2430.

[47] Liqun Qi. 1984. Some simple estimates for singular values of a matrix. *Linear Algebra and Its Applications* 56 (1984), 105–119.

[48] Nadine Renard, Salah Bourennane, and Jacques Blanc-Talon. 2008. Denoising and dimensionality reduction using multilinear tools for hyperspectral images. *IEEE Geoscience and Remote Sensing Letters* 5, 2 (2008), 138–142.

[49] Christian Rogass, Christian Mielke, Daniel Scheffler, Nina Boesche, Angela Lausch, Christin Lubitz, Maximilian Brell, Daniel Spengler, Andreas Eisele, Karl Segl, et al. 2014. Reduction of uncorrelated striping noise—applications for hyperspectral pushbroom acquisitions. *Remote Sensing* 6, 11 (2014), 11082–11106.

[50] Sameera V Mohd Sagheer and Sudhish N George. 2019. Denoising of low-dose CT images via low-rank tensor modeling and total variation regularization. *Artificial intelligence in medicine* 94 (2019), 1–17.

[51] Feng Shi, Jian Cheng, Li Wang, Pew-Thian Yap, and Dinggang Shen. 2015. LRTV: MR image super resolution with low rank and total variation regularizations. *IEEE transactions on medical imaging* 34, 12 (2015), 2459–2466.

[52] Arizona State University. [n.d.]. *Yuv video sequences." [Online]. Available:*. http://trace.eas.asu.edu/yuv/

[53] Tong Wu and Waheed U Bajwa. 2018. A Low Tensor-Rank Representation Approach for Clustering of Imaging Data. *IEEE Signal Processing Letters* 25, 8 (2018), 1196–1200.

[54] Luofeng Xie, Ming Yin, Xiangyun Yin, Yun Liu, and Guofu Yin. 2018. Low-rank sparse preserving projections for dimensionality reduction. *IEEE Transactions on Image Processing* 27, 11 (2018), 5261–5274.

[55] Yang Xu, Zebin Wu, Jocelyn Chanussot, and Zhihui Wei. 2018. Joint reconstruction and anomaly detection from compressive hyperspectral images using mahalanobis distance-regularized tensor RPCA. *IEEE Transactions on Geoscience and Remote Sensing* 56, 5 (2018), 2919–2930.

[56] Qiangqiang Yuan, Liangpei Zhang, and Huanfeng Shen. 2013. Hyperspectral image denoising with a spatial–spectral view fusion strategy. *IEEE Transactions on Geoscience and Remote Sensing* 52, 5 (2013), 2314–2325.

[57] Linwei Yue, Huanfeng Shen, Jie Li, Qiangqiang Yuan, Hongyan Zhang, and Liangpei Zhang. 2016. Image super-resolution: The techniques, applications, and future. *Signal Processing* 128 (2016), 389–408.

[58] Hongyan Zhang, Wei He, Liangpei Zhang, Huanfeng Shen, and Qiangqiang Yuan. 2013. Hyperspectral image restoration using low-rank matrix recovery. *IEEE Transactions on Geoscience and Remote Sensing* 52, 8 (2013), 4729–4743.

[59] Zemin Zhang and Shuchin Aeron. 2015. Denoising and completion of 3D data via multidimensional dictionary learning. *arXiv preprint arXiv:1512.09227* (2015).

[60] Zemin Zhang and Shuchin Aeron. 2016. Exact tensor completion using t-SVD. *IEEE Transactions on Signal Processing* 65, 6 (2016), 1511–1526.

[61] Zemin Zhang, Gregory Ely, Shuchin Aeron, Ning Hao, and Misha Kilmer. 2014. Novel methods for multilinear data completion and de-noising based on tensor-SVD. In *Proceedings of the IEEE conference on computer vision and pattern recognition*. 3842–3849.

[62] Yong-Qiang Zhao and Jingxiang Yang. 2014. Hyperspectral image denoising via sparse representation and low-rank constraint. *IEEE Transactions on Geoscience and Remote Sensing* 53, 1 (2014), 296–308.
[63] Pan Zhou and Jiashi Feng. 2017. Outlier-robust tensor PCA. In *Proceedings of the IEEE Conference on Computer Vision and Pattern Recognition*. 2263–2271.
[64] Xiaowei Zhou, Can Yang, Hongyu Zhao, and Weichuan Yu. 2015. Low-rank modeling and its applications in image analysis. *ACM Computing Surveys (CSUR)* 47, 2 (2015), 36.